\documentclass{article}

\usepackage{arxiv}

\usepackage[utf8]{inputenc} 
\usepackage[T1]{fontenc}    
\usepackage{hyperref}       
\usepackage{url}            
\usepackage{booktabs}       
\usepackage{amsfonts}       
\usepackage{nicefrac}       
\usepackage{microtype}      
\usepackage{lipsum}
\usepackage{graphicx}
\graphicspath{{./figures/}}
\DeclareGraphicsExtensions{.pdf}
\usepackage[cmex10]{amsmath}

\title{3DCAM: A Low Overhead Crosstalk Avoidance Mechanism for TSV-Based 3D ICs}

\author{
  Reza Mirosanlou \thanks{This work has been done while the authors were at Sharif University of Technology.} \\
  University of Waterloo\\
  \texttt{rmirosan@uwaterloo.ca} \\
   \And
  Mohammadkazem Taram \\
  University of California San Diego\\
  \texttt{mtaram@cs.ucsd.edu} \\
  \AND
  Zahra Shirmohammadi \\
  Shahid Rajaee Teacher Training University\\
  \texttt{shirmohammadi@kish.sharif.edu} \\
  \AND
  Seyed-Ghassem Miremadi \\
  Sharif University of Technology\\
  \texttt{miremadi@sharif.edu}\\
}

\begin{document}
\maketitle

\begin{abstract}
\emph{Three Dimensional Integrated Circuits} (3D IC) offer lower power consumption, higher performance, higher bandwidth, and scalability over the conventional two dimensional ICs. \emph{Through-Silicon Via} (TSV) is one of the fabrication mechanisms that connects stacked dies to each other. The large size of TSVs and the proximity between them lead to undesirable coupling capacitance. This interference causes mutual influences between adjacent TSVs and produces crosstalk noise. Furthermore, this effect threats the reliability of data during traversal between layers. This paper proposes a mechanism that efficiently reduces crosstalk noise between TSVs with lower area overhead as compared to previous works. This mechanism revolves around the fact that retaining TSV value in current state can reduce coupling in some cases.
 To evaluate the mechanism, gem5 simulator is used for data extraction and several benchmarks are taken from the SPEC2006 suite. The simulation results show that the proposed mechanism reduces crosstalk noise with only 30\% imposed TSV overhead while delay decreased up to 25.7\% as compared to a recent related work.
\end{abstract}


\section{Introduction}

Technology node scaling in recent decades ushered in gate delay cut-off and rise of interconnection latency \cite{Agarwal2006}. Hence, interconnects have become a major performance bottleneck of high performance \emph{System-on-Chips} (SoC) and \emph{Integrated Circuits} (IC) \cite{Pande2007}. In addition, interconnections have become more susceptible to noises in particular crosstalk \cite{Pande2007}. 
On the other hand, the advent of multi-core processors with ever increasing number of cores has highlighted the need for fast and reliable interconnections. 
One of the potential solutions to alleviate the interconnection delay problem is the three dimensional integration using \emph{Through-Silicon Vias} (TSV). Vertical integration of IC dies using TSVs offers high density connections between adjacent dies. This technology also allows stacking of dies with nonidentical technologies such as CMOS with high density DRAM which can be used as a solution to mitigate memory wall problem \cite{Loh}. Furthermore, the average and maximum distance between interconnect nodes of the 3D stacked ICs are greatly decreased which leads to significant delay, power, and area improvement. Despite of the TSV advantages, the adjacent, short and bounded TSVs are prone to TSV-to-TSV coupling and crosstalk noise which increases transmission time and power consumption, and more importantly, it threats the signal integrity \cite{Liu2011,Kumar2013}.

\begin{figure}[!t]
\centering
\includegraphics[width=3.57in]{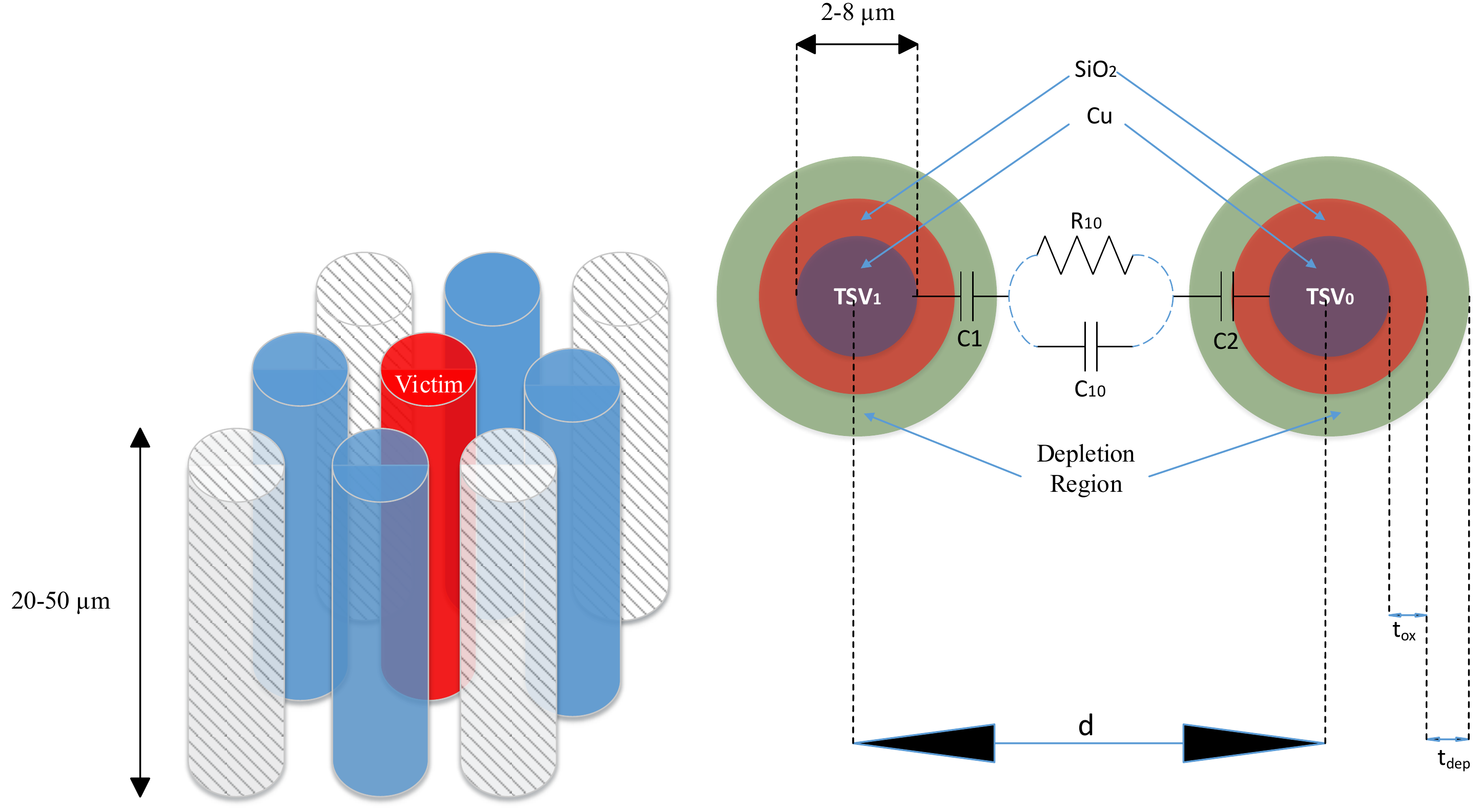}
\centering

\caption{Overview of coupling characteristics in TSVs \cite{Engin2013}; According to ITRS \cite{itritrs}, it is predicted that the height of TSVs will reach to  20-50 $\mu m$ and the via diameter will be 2-8 $\mu m $  till 2018. }
\label{first}
\end{figure}

As demonstrated in Fig.~\ref{first}, every TSV may be surrounded by neighbour TSVs which cause a big and undesirable coupling noise. This TSV-to-TSV coupling could be very challenging in 3D ICs due to the fact that TSVs are large and thick, thus the coupling between two adjacent TSVs can be huge.  Moreover, the effective coupling capacitance between TSVs doubles when the aggressor and the victim signals switch in opposite directions\cite{Liu2011}.

Plenty of crosstalk minimization methods have been proposed in the literature of 2D design (e.g. \cite{ganguly2008design,Duan2009,Ganguly2009very,Zhang2004,Hirose2000}). However, these methods cannot be directly applied to alleviate TSV-to-TSV crosstalk noise,  inasmuch as the TSVs are not placed in the same planar and are greatly affected by more than two aggressors \cite{Zou2014}. 
Recent efforts in TSV-to-TSV crosstalk minimization including \cite{Zou2014,Kumar2013,Chang2013} are complex and impose significant area and TSV overhead. SheildUS \cite{Chang2013}, by adding a crossbar, remaps data to TSVs in order to shield more active signals by the signals which predicted to have less transitions in the future. In addition to its complex decision-making circuit, the accuracy of its predictor is under question due to the fact that the signals may not have a regular pattern.
3DLAT \cite{Zou2014} exploits less adjacent codes to limit maximum number of transitions in adjacent TSVs. \emph{Crosstalk Avoidance Codes} (CAC) \cite{Kumar2013} is another coding scheme for TSV-to-TSV crosstalk minimization. These approaches also need a complex and large coder and also suffer from a considerable information redundancy overhead.

In this paper, we propose a TSV-to-TSV crosstalk minimization method, named 3DCAM, which can effectively reduce coupling noise between TSVs with a relatively low area and TSV overhead. In addition, the proposed method uses a small simple coder which reduces run-time performance overhead. In the case of a transition on a target signal, considering the target's neighbours and their coupling effect, 3DCAM decides to whether retain target's value or send its original transition. In the condition that coder decides to retain the value it informs the decoder through a control TSV. The simulation results show that 3DCAM can reduce the transmission delay up to 25.7\% as compared to 3DLAT mechanism. 3DCAM imposes only 30\% TSV overhead which is much less than the 3DLAT TSV overhead (which is 80\% for $\omega=4$).  

The rest of this paper is structured as follows. In Section II, related works are reviewed. Section III describes the crosstalk model for TSVs based 3D ICs on which 3DCAM is built. In Section IV we present 3DCAM crosstalk avoidance mechanism. Section V explains the simulations and results and, finally, Section VI concludes the paper.

\begin{figure}[!t]
\centering
\includegraphics[width=2in]{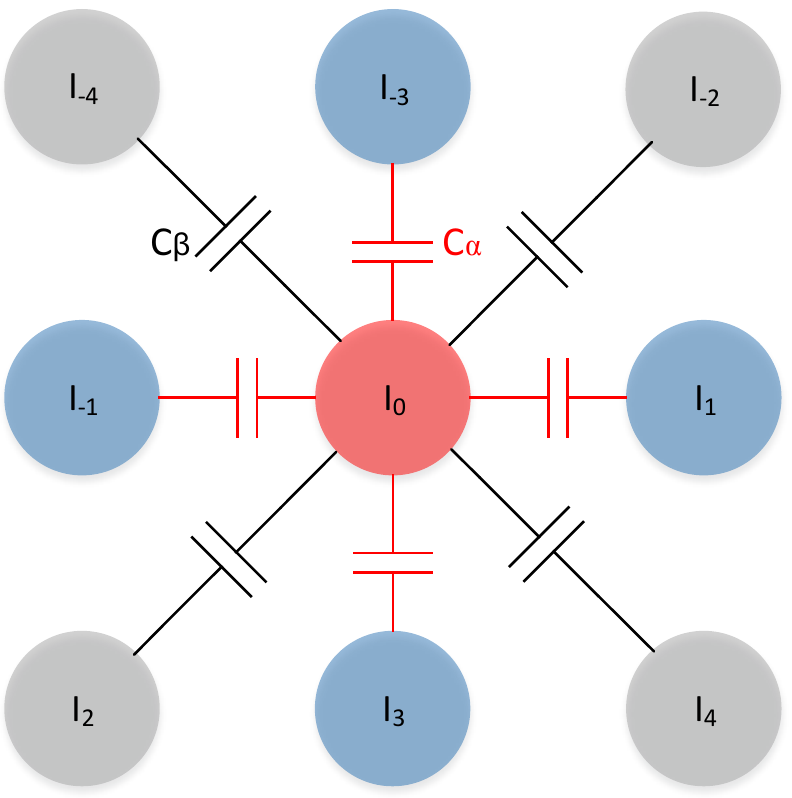}
\centering
\caption{The coupling capacitance crosstalk model for $3\times3$ TSV cluster}
\label{model}
\end{figure}

\section{Related Work} \label{previous}

In the context of \emph{Two Dimensional Network-on-Chips} (2D NoC), there are plenty of works that target power consumption~\cite{performance,toot,smart}, reliability~\cite{performance}, security~\cite{securitry}, or performance~\cite{performance} of the interconnections. Particularly, crosstalk minimization methods can be classified in three categories: physical level, transistor level  and, \emph{Register Transfer Level} (RTL) techniques. Wire spacing \cite{Agarwal2006}, active and passive shielding \cite{Zhang2004}, and buffer insertion \cite{Akl2008} are examples of physical level techniques. \cite{Hirose2000} is a transistor level mechanism which reduces the crosstalk noise by skewing the simultaneous opposite transitions. Although this approach reduces the crosstalk, it requires timing adjustment between senders and receivers. Furthermore, this approach suffers from run-time management. 
The general idea behind RTL level techniques is to omit some undesirable transition patterns by using coding schemes. There are variety of works that focused on analytical aspect and coding concepts \cite{Duan2009,sridhara2007coding}. Error detection codes and error correction codes \cite{ganguly2008design}, joint crosstalk avoidance mechanism \cite{sridhara2007coding}, and CACs \cite{ganguly2008design} are examples of these coding schemes. 

Although the above approaches may cope with crosstalk in 2D ICs, they cannot be directly applied in 3D technologies because the additional dimension makes consequential differences in crosstalk problem analysis. Gathering the long and thick TSVs causes new reliability issues which have been studied recently \cite{Okoro2014,Jung2014}. Several mechanisms have been proposed to make 3D ICs more reliable against crosstalk noise, e.g., \cite{Kumar2013,Eghbal2014,Zou2014,Chang2013,Eghbal2015}.    
The TSV-to-TSV capacitance and inductance coupling are two major threats to 3D IC reliability. Previous works have concentrated on these effects from two perspectives. \cite{Kumar2013,Zou2014,Chang2013} proposed capacitance-based mechanisms and \cite{Eghbal2014,Eghbal2015,Motoyoshi2009} proposed inductance-based techniques to reduce crosstalk effects in 3D ICs. 

Increasing TSV distances from each other, shielding TSVs, inserting buffers at the victim side, inserting buffers, decreasing driver size at the aggressor side, and increasing load at the wires are the mechanisms examined in \cite{Liu2011} to mitigate TSVs crosstalk noise. According to their experiments, unlike 2D wires, increasing TSV distances is not an effective solution to TSV-to-TSV coupling problem and the other solutions either need high effort at post-design time or have negative impact on timing performance. 

RTL mechanisms in 3D IC have been proposed and experimented recently. \cite{Kumar2013} proposed a coding scheme that reduces the maximum crosstalk about 28\% based on their proposed crosstalk model. Two other mentionable mechanisms in 3D IC against crosstalk noise which have been studied recently are ShieldUS \cite{Chang2013} and 3DLAT \cite{Zou2014}. ShieldUS uses data with less transitions as the shield for the more active data. SheildUS tries to minimize average transmission time with run-time mechanism that remaps data to TSVs in order to banish links with specific crosstalk pattern from others. The TSV overhead of this method is not considerable because it uses the same data as shield. But a large crossbar is required for bit shuffling which imposes considerable area overhead. This crossbar will get larger by increasing the number of bits to shuffle. Besides, this method needs the data to be highly regular, as this method tries to predict the activity of the signals.

The authors in \cite{Zou2014} introduce use of less adjacent transition code along with transition signaling to minimize the number of transitions. Furthermore, 3DLAT reduces higher crosstalk class frequency. This scheme has a significant TSV overhead which is not negligible. According to the authors' report, TSV overhead of 3DLAT is about 80\% with $\omega=4$. This mechanism imposes more than 160\% area overhead with $\omega=2$ for the same bitwidth. This method also suffers from significant area overhead which imposed by its complex coder.

\begin{figure}[!t]
\centering
\includegraphics[width=3in]{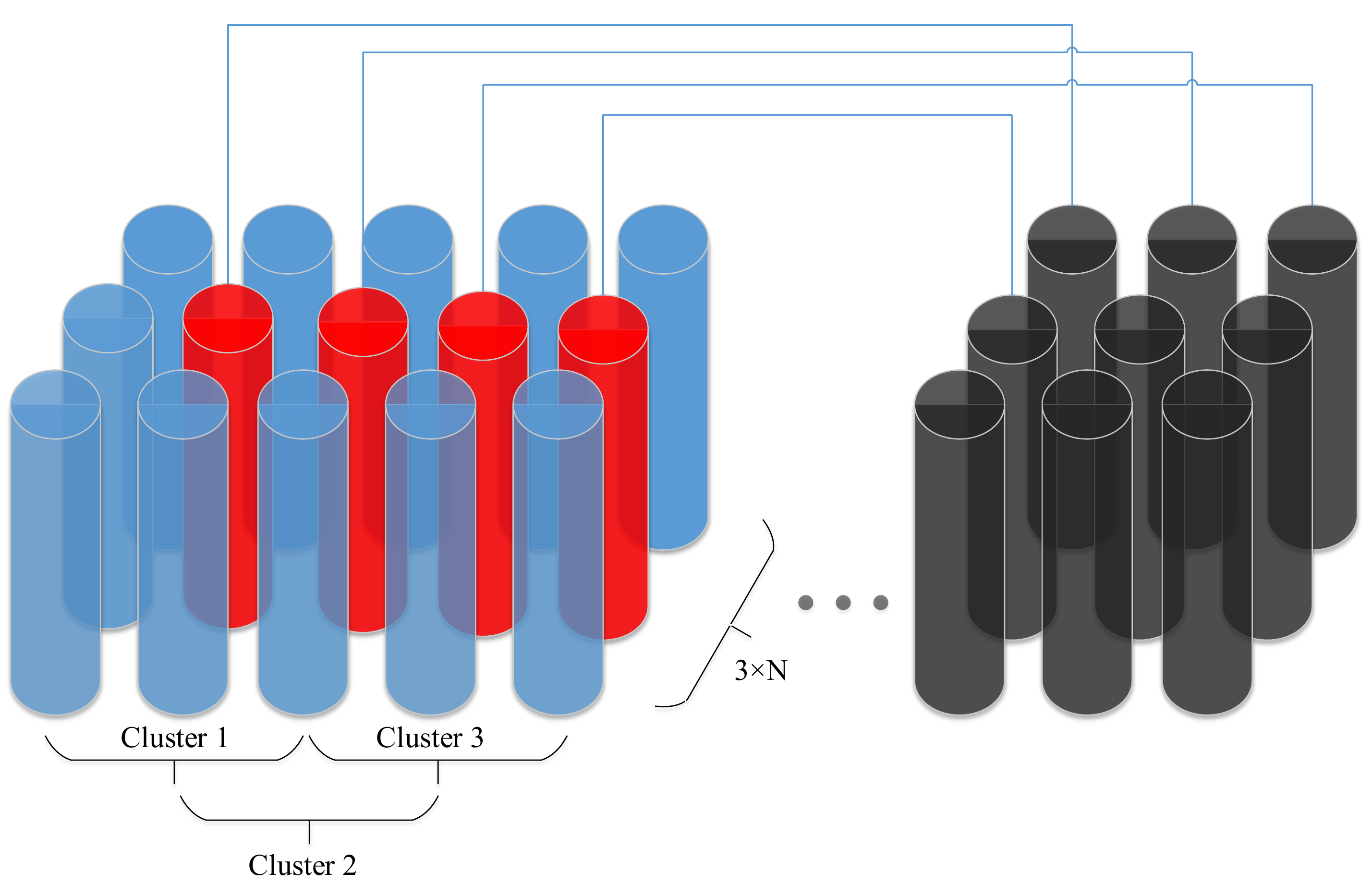}
\centering
\caption{Layout of control TSVs for $3\times N$ bus}
\label{control}
\label{layout}
\end{figure}
\section{3D Crosstalk Coupling Noise Model} \label{modelsection}
In 2D integrated circuits, three neighbor wires affect each other and create coupling capacitance. The effective coupling capacitance which imposed on the victim (i.e. middle) wire is modeled by Eq.~(\ref{equation.1}) \cite{Hirose2000}. 

\begin{equation} \label{equation.1}
\begin{split}
C_{eff}=C_{G} + C_{C}\mid\frac{\Delta V - \Delta V_{-1}}{V_{dd}}\mid \\
+ C_{C} \mid \frac{\Delta V - \Delta V_{+1}}{V_{dd}}\mid
\end{split}
\end{equation}

Where $\Delta V$ is swing voltage on victim wire, $\Delta V_{-1}$ and $\Delta V_{+1}$ are the voltages that switch on neighbor wires and $V_{dd}$ is the supply voltage. In addition, $C_{c}$ is coupling capacitance that is imposed between the victim wire and its neighbors and, $C_{G}$ is the coupling capacitance between substrate and plate.

Based on the Eq.~(\ref{equation.1}), we can model the transmission delay by the Eq.~(\ref{equation.3}):  

\begin{equation}  \label{equation.3}
	\begin{split}
	 \tau = (1+\rho \lambda)\pi_{o}
	\end{split}
\end{equation}

Where $\rho$ is the coupling coefficient of adjacent wires, $\lambda$ is the coupling capacitance to substrate capacitance ratio ($C_{C} /C_G  $) and $\pi_{0}$ is the delay of a wire in the ideal channel, i.e., a channel without any coupling effect such as capacitance and inductance. For instance, when the both aggressors and the victim wire, switch in opposite directions a delay equal to $(1+4\lambda)\pi_{0}$ will be imposed to the channel. 

Similar to the 2D IC crosstalk model, we can drive a delay model for TSVs. Akin to previous studies \cite{Kumar2013,Chang2013} and based on TSV's inherent properties, a square model of 9 adjacent TSVs is considered. Fig.~\ref{model} depicts 9 neighbor TSVs from top view. We discuss the crosstalk effect on the victim TSV (specified by red in Fig.~\ref{model}) and we model coupling capacitance noise that is emanated from its neighbor TSVs. Direct neighbors (i.e. north, south, west, and east) are closer to the victim, and thus their coupling capacitances are more destructive than the coupling effects of diagonal neighbors (i.e. northeast, northwest, southeast, and southwest). In order to model the effective capacitance on the victim TSV, we can extend the Eq.~(\ref{equation.1}) to Eq.~(\ref{equation.4}):
\begin{equation}  \label{equation.4}
	\begin{split}
	C_{eff}=C_{G} &+ \sum_{i=-2}^{2} C_{\alpha} {  \mid\frac{\Delta V_{I_{0}} - \Delta V_{I_{2i}}}{V_{dd}}\mid} \\
	\\
	&+ \sum_{i=-1}^{2} C_{\beta}  { \mid\frac{\Delta V_{I_{0}} - \Delta V_{I_{2i-1}}}{V_{dd}}\mid}
	\end{split}
\end{equation}

Where  $C_{\alpha}$ represents coupling capacitance between a direct aggressor and the victim and $C_{\beta}$ is coupling capacitance between diagonal aggressor
and the victim. 
Similar to 2D crosstalk delay model, based on Eq.~(\ref{equation.4}) we can extend the Eq.~(\ref{equation.3}) to model 3D TSVs as follows:

\begin{equation}  \label{equation.5}
	\begin{split}
	\tau = (1+\rho_1 \lambda_1+ \rho_2 \lambda_2)\pi_{o}
	\end{split}
\end{equation}

Where $\rho_1 $ is the coupling coefficient of the direct aggressors, $\lambda_1$ is the direct coupling capacitance to substrate capacitance ratio ($C_{\alpha} /C_G  $), $\rho_2 $ is the coupling coefficient of the diagonal aggressors, and $\lambda_2$ is the diagonal coupling capacitance to substrate capacitance ratio ($C_{\beta} /C_G  $). $\rho_1 $ and $\rho_2 $ indicate how the changes in aggressors voltages affect the crosstalk on the victim. For instance, if a direct neighbor switches in opposite direction,  $\rho_1 $ would increase by two. In the worst case, all the neighbours switch from $V_{dd}$ to zero and the victim switches from zero to $V_{dd}$. In this case the transmission delay would be $(1+ 8 \lambda_1+ 8 \lambda_2)\pi_{o}$.
\begin{equation}  \label{equation.2}	
	\begin{split}
	C_{eff}  &= [(C_{\alpha} + C_{\beta}) \times 8 ]+ C_{G} \\
			 &= [1.5 C_{\beta} + C_{\beta} \times 8] + C_{G} \\
			 &= 20 \times C_{\beta} + C_{G}
	\end{split}
\end{equation}

As reported in \cite{Zou2014,Chang2013} the  $ \lambda_1 =  5.54 $ and  $ \lambda_2 =  3.92 $. For the sake of simplicity, we assume $ \lambda_1 = 1.5 \lambda_2 $ and consequently $C_\alpha = 1.5 C_\beta$, thus as Eq.~(\ref{equation.2}) shows, we can classify crosstalk patterns in 40 distinct classes which represented in Table.~\ref{table.1}. Indeed, patterns with higher class numbers have higher crosstalk noise and delay.

\begin{table}
\renewcommand{\arraystretch}{2}
\caption{TSV Crosstalk Classes}
\label{table.1}
\centering
\begin{tabular}{| c| c| c|}
\hline
Class & $ C_eff $ & $T_{i-4,...,i+4}$(t)$\to$$T_{i-4,...,i+4}$(t+1)\\
\hline
0 & $C_G$ & 000000000 $\to$ 111111111\\
1 & $C_G+C_\beta$ & 000000000 $\to$ 011111111\\
2 & $C_G+1.5C_\beta$ & 100000000 $\to$ 011111111\\
3 & $C_G+2C_\beta$ & 010000000 $\to$ 101111111\\
\textbf{.} &\textbf{.} & \textbf{.} \\
\textbf{.} &\textbf{.} & \textbf{.} \\
\textbf{.} &\textbf{.} & \textbf{.} \\
39 & $C_G+20C_\beta$ & 000010000 $\to$ 111101111\\
\hline
\end{tabular}
\end{table}

\section{Proposed Mechanism} \label{proposed}

\subsection{Overview}
  As mentioned in Section~\ref{modelsection}, we have classified the patterns into 40 different classes based on their crosstalk effects. In this section, we propose 3DCAM mechanism which aims to minimize crosstalk effect on signal integrity and performance. To this end, we need to minimize the occurrences of higher crosstalk classes and maximize the occurrences of lower crosstalk classes. To accomplish which, we have to change the transition patterns on TSVs. As we know about the transmission line, one of the following conditions can occur to a single wire: $0 \rightarrow 1$, $1 \rightarrow 0$, $0 \rightarrow 0$, or $1 \rightarrow 1$. 
The motivation behind this work is the fact that retaining the previous value of a victim TSV can significantly affect the transmission's crosstalk class. 

Table.~\ref{table.2} presents some examples, in which retaining the victim (middle) TSVs significantly reduces crosstalk class. The first column of this table is the incoming pattern in which the victim TSV could have transition and the second column shows the same pattern except that the victim's transition is eliminated. The up and down arrows show zero to $V_{dd}$ and  $V_{dd}$ to zero transitions respectively, and dashes(\lq-\rq) ~  denote no transition on TSVs. The changes in crosstalk classes are represented in third column of this table. For instance, the first row of Table. \ref{table.2} shows the case that falls into crosstalk class of $24C$ because the victim has three direct neighbors with opposite directions ($C_{eff}$ increased by $3 \times 2 \times 1.5 C_\beta$), a direct neighbor without transition ($C_{eff}$ increased by $1 \times 1 \times 1.5 C_\beta$), two diagonal neighbors with same direction transitions (impose no crosstalk) and two diagonal neighbors with no transitions ($C_{eff}$ increased by $2 \times 1 \times 1 C_\beta$). So the $C_{eff}$ will be $C_G + 12.5C_\beta$ and according to Table.\ref{table.1} the pattern falls into 24C class. By similar manner the crosstalk class after eliminating the victim's transition would be 12C.
As depicts in this table, the crosstalk minimization achieved by this simple modification is considerable.

\begin{table}[!h]
\caption{Motivational Example}
\label{table.2}
\includegraphics[width=2.5in]{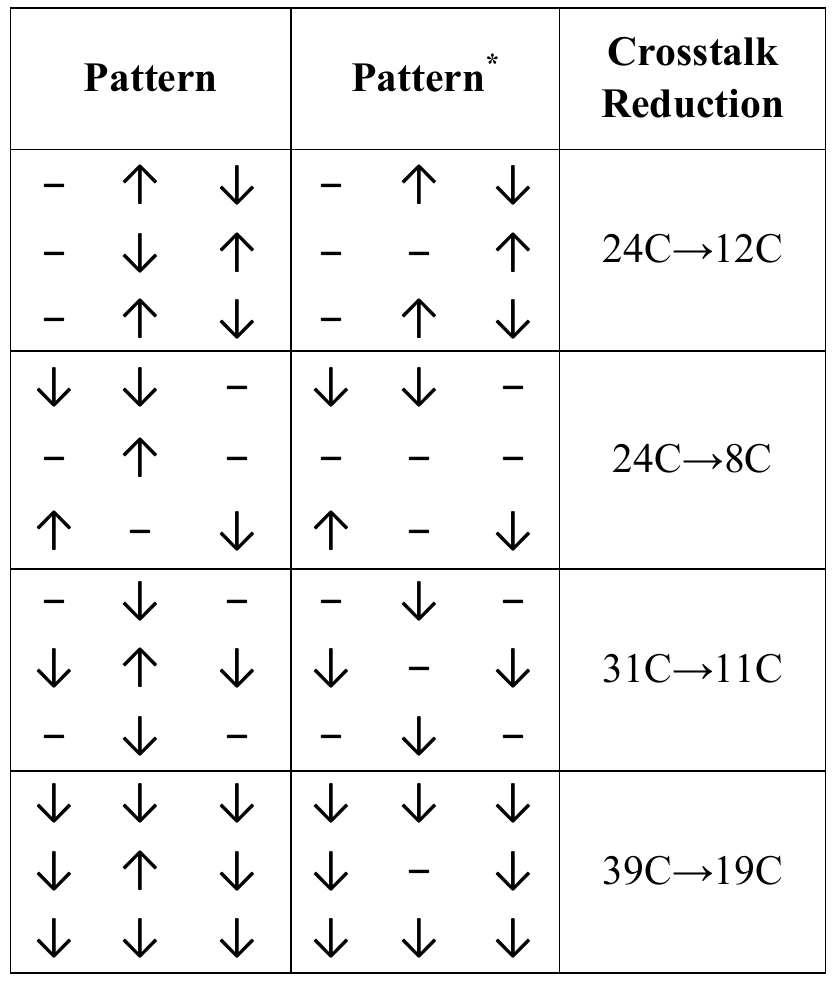}
\centering

\end{table}
\subsection{Control TSVs}
In order to retain victim TSV's value, 3DCAM should by some means inform the receiver side decoder that the victim's transition has been eliminated.  
 Thus, 3DCAM reserves some TSVs for this purpose. These control TSVs only switch when the value of their corresponding victim TSVs are not valid (i.e. the 3DCAM eliminates their transitions). Fig.~\ref{control} demonstrates the layout of these control TSVs for a $3 \times N$  link. The middle TSVs of every $3\times3$ cluster (including overlapping ones) have a control TSV through which 3DCAM coder informs decoder that the value of the TSV is valid or not. Since the control TSVs may have coupling between themselves, we can repeat the technique and apply 3DCAM on them.

\subsection{Switch Threshold} \label{thresholdsection}
Retaining the victim's value is not always beneficial to crosstalk class. In the cases that the original transitions are good enough, retaining the victim's value may result in a worse crosstalk class. In addition, it has negative impact on control TSVs, considering the fact that retaining a value requires a transition in a control signal. Table.~\ref{table.3} shows an example in which retaining the victim's value leads to a worse crosstalk class and hence a worse transmission delay.

\begin{table}[!h]
\caption{Disruptive Retaining Example}
\label{table.3}
\includegraphics[width=2.5in]{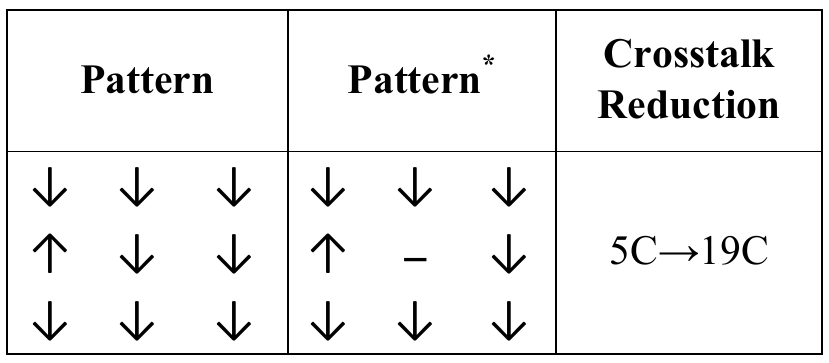}
\centering
\end{table}

To address this issue, we introduce a parameter called \emph{Switch Threshold} (ST). This parameter determines which patterns should be manipulated by 3DCAM. In the other words, 3DCAM retains the value of the victim TSV, only if the transitions of neighbour TSVs make a pattern which has a crosstalk class more than ST. 

\begin{figure}[!t]
\centering
\includegraphics[width=3.5in]{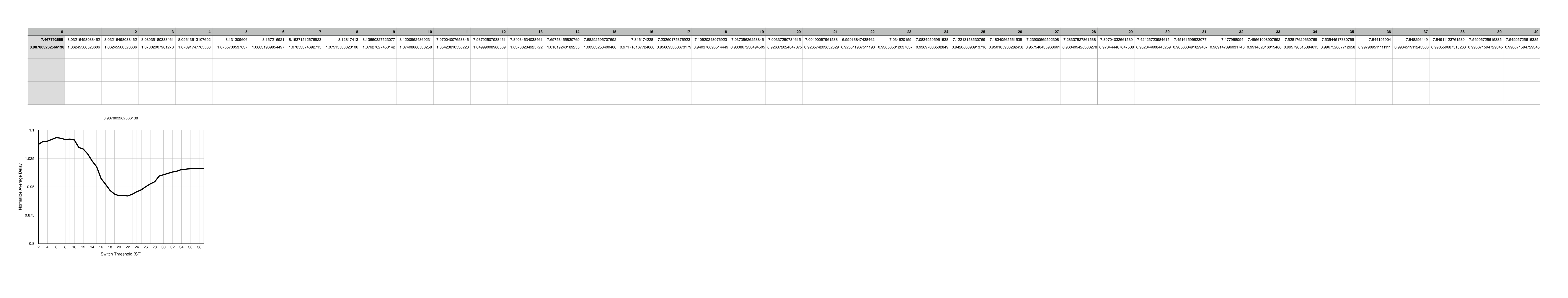}
\centering
\caption{Improvement of 3DCAM for different values of the threshold parameter}
\label{threshold}
\end{figure}

To find the optimal value for the ST parameter, we swept ST parameter from 0C to 39C and measured average transmission delay of several benchmarks. Fig. \ref{threshold} depicts the results of this experiment. As this figure shows, the average delay is minimum when the ST is set to 20. These results are also consistent with the intuition, seeing that setting ST to 20 bisects the crosstalk classes.

\subsection{CODEC Design}
\begin{figure}[!t]
\centering
\includegraphics[width=3.2in]{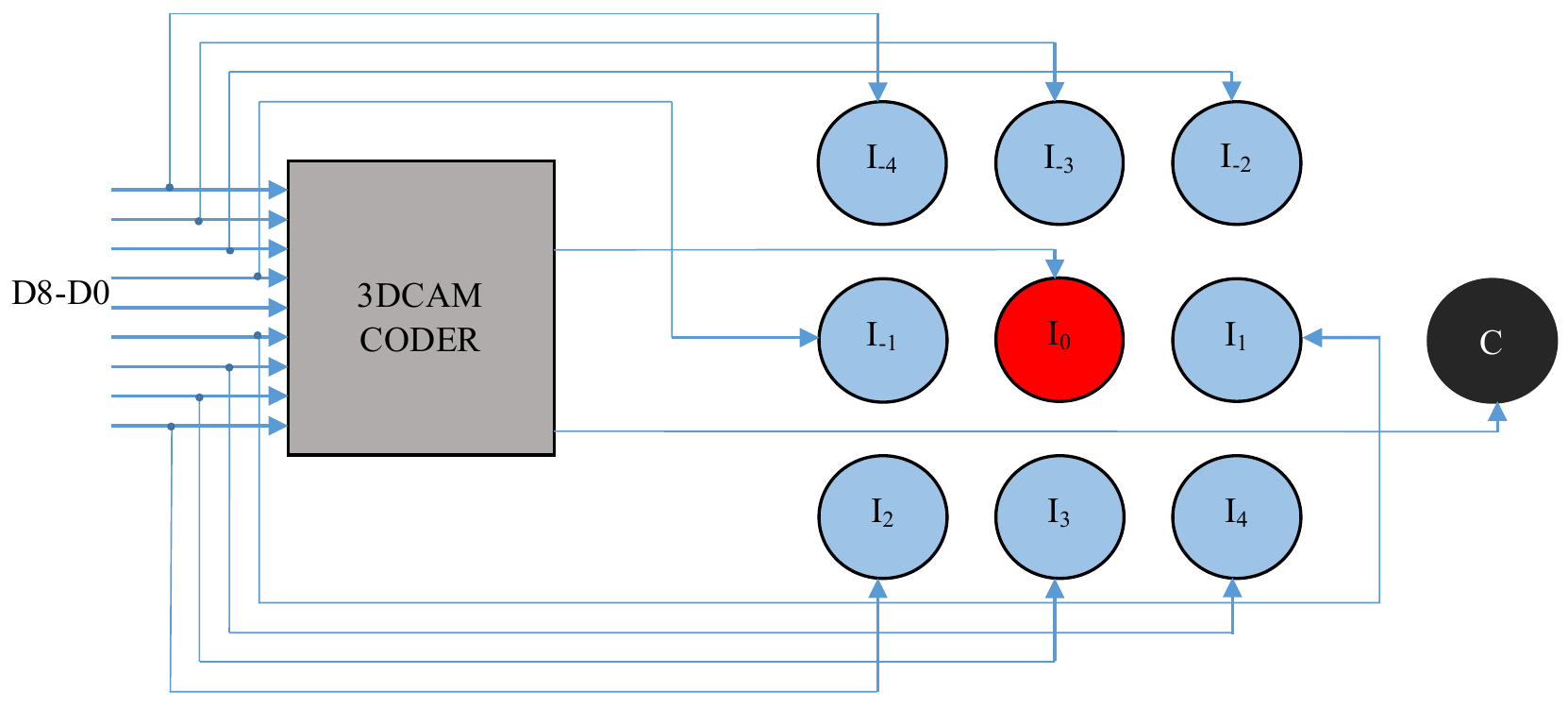}
\centering
\caption{3DCAM coding mechanism}
\label{coder}
\end{figure}

\begin{figure}[!t]
\centering
\includegraphics[width=3.2in]{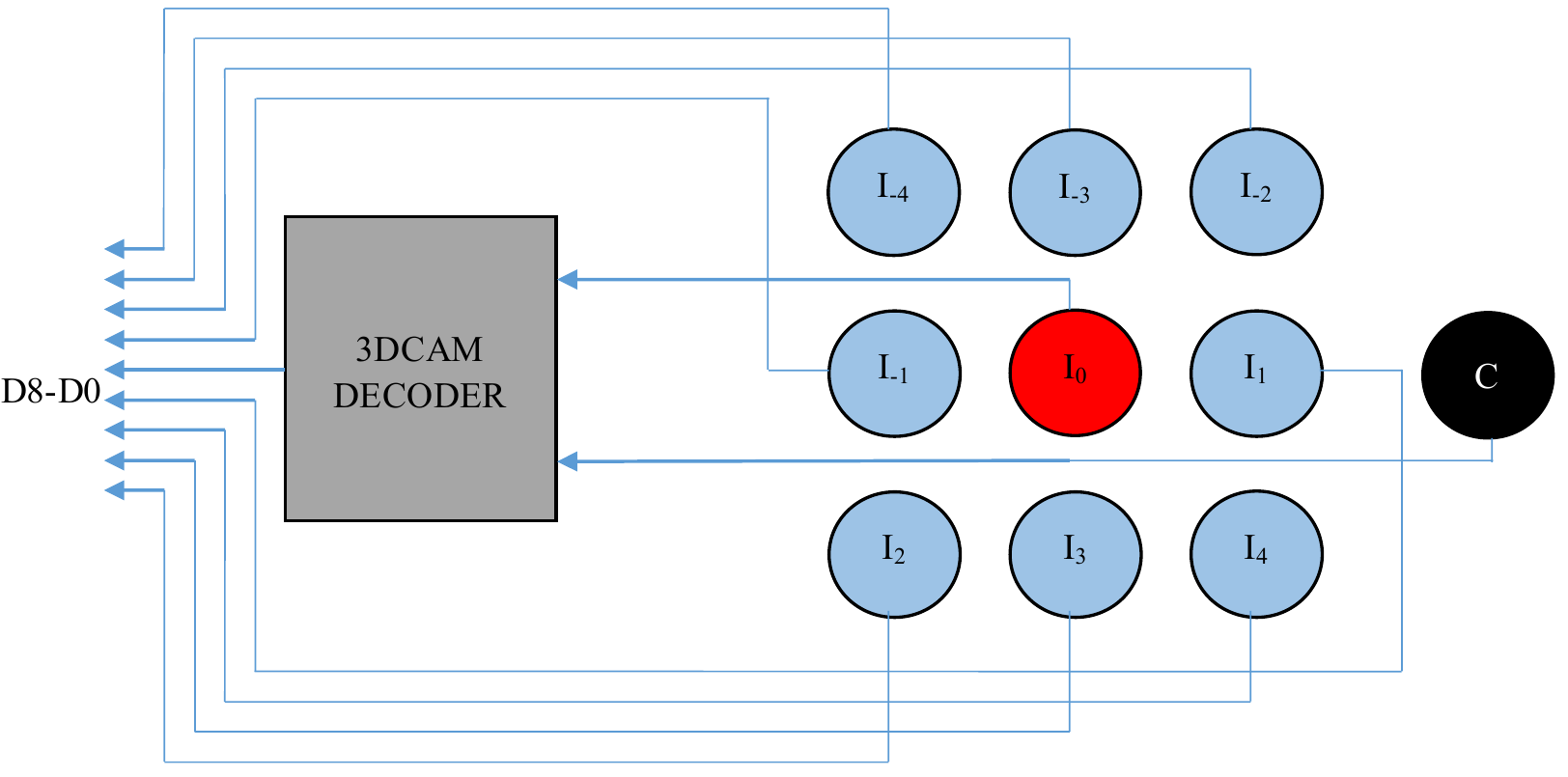}
\centering
\caption{3DCAM decoding mechanism}
\label{decoder}
\end{figure}

Fig. \ref{coder} and Fig. \ref{decoder} illustrate the structure of 3DCAM coder and decoder. 
In order to send 9 bits of data ($D_{0}$ - $D_{8}$) from die $X$ to die $Y$, the data have to be delivered to the coder which has been placed in die $X$. After that, the coder evaluates the crosstalk class which will be imposed to the victim TSV ($I_0$). Then it checks whether the crosstalk class is greater than the ST parameter, in which case, the coder eliminates the victim's transition and flips the control bit. Next, the data and control bit are sent to the die $Y$ through TSVs. At the die $Y$, decoder receives the data and control signals. Then based on the control signal, the decoder determines the original value of the victim TSV. It is noteworthy that due to straightforward functionality of the 3DCAM coder and decoder, they can be implemented by simple circuits. 

\subsection{TSV Overhead} \label{TSVoverhead}
\begin{figure}[!t]
	\centering
	\includegraphics[width=3.5in]{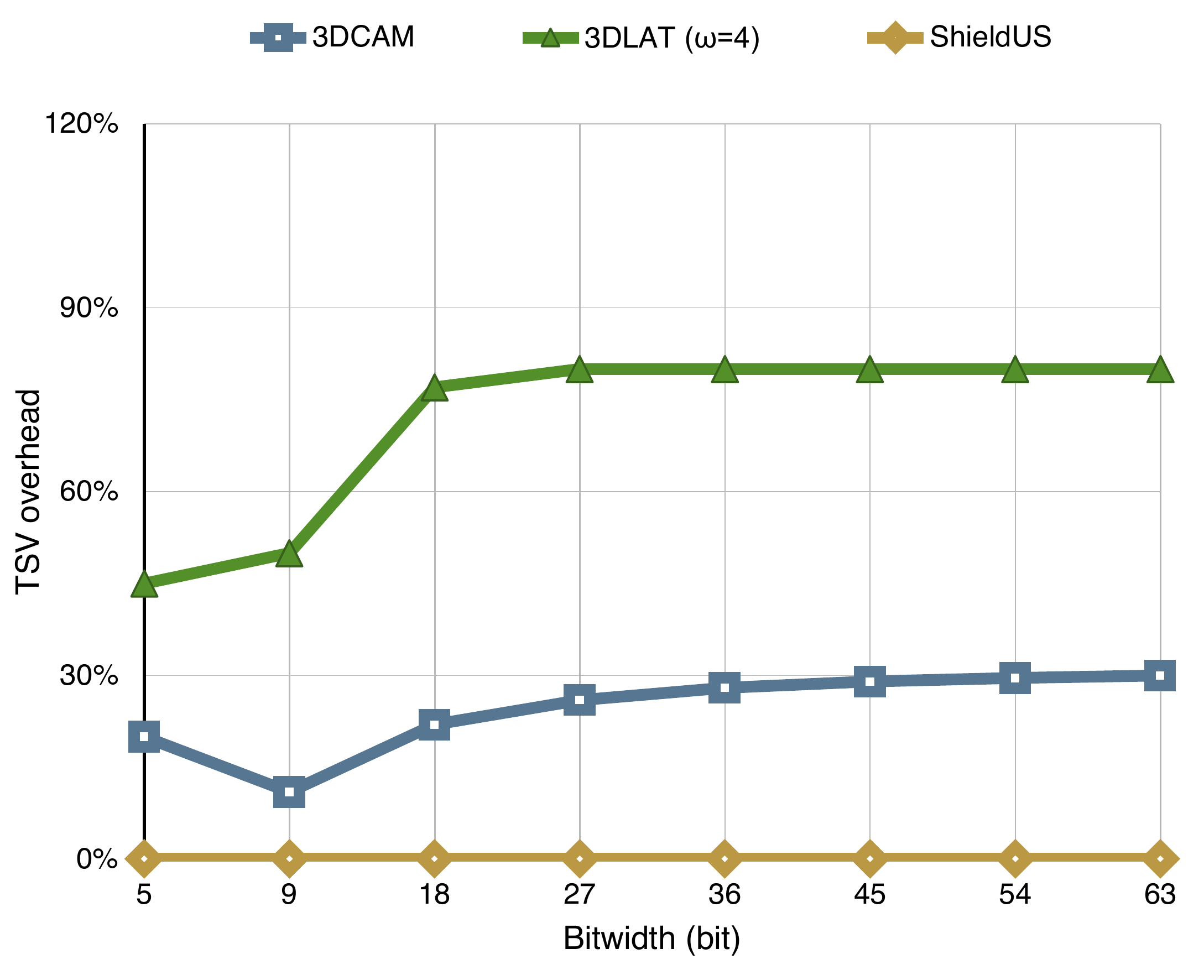}
	\centering
	\caption{percentage of extra TSVs needed for 3DCAM, 3DLAT, and ShieldUS mechanisms}
	\label{percentage}
	\end{figure}

Because of the controlling mechanism, we have to reserve an extra TSV for each cluster including 9 TSVs. As a result, 3DCAM suffers from about 30\% TSV overhead. Fig.~\ref{percentage} demonstrates the TSV overhead of proposed method, ShieldUS, and 3DLAT $(\omega=4)$. As Fig.~\ref{percentage} depicts, the TSV overhead of 3DLAT is about 80\% and ShieldUS imposed no TSV overhead to the circuit because it only shuffles and remaps the data. However, ShieldUS needs a considerable $9 \times 9$ crossbar which is used to remap data to TSVs. 

\section{Evaluation} \label{evaluation}
In this section, the proposed mechanism is evaluated and compared with two 3D crosstalk reduction schemes, 3DLAT \cite{Zou2014} (with $\omega=4$) and ShieldUS \cite{Chang2013} (with $interval=100$). Based on the crosstalk model that proposed in Section \ref{modelsection}, we measured the amount of crosstalk  reduction and average delay on real traces which are taken from SPEC2006 benchmark suite \cite{spec}. We used gem5 simulator \cite{gem5} to capture the transitions of memory data bus of \textit{gcc, mcf, namd, soplex, h264, omnetpp, xalancbmk, perlbench2, bwaves, cactusADM, dealII, lbm}, and \textit{aster} benchmarks. 

Without loss of generality, we assume that the TSVs are arranged in $3 \times N$ layout. Also, we suppose that the data bitwidth is 64 and thus we need eight  $3 \times 3 $ TSV clusters for the data and three clusters for control lines.

\subsection{Delay Analyisis}

\begin{figure*}[!t]
	\centering
	\includegraphics[width=5.5 in]{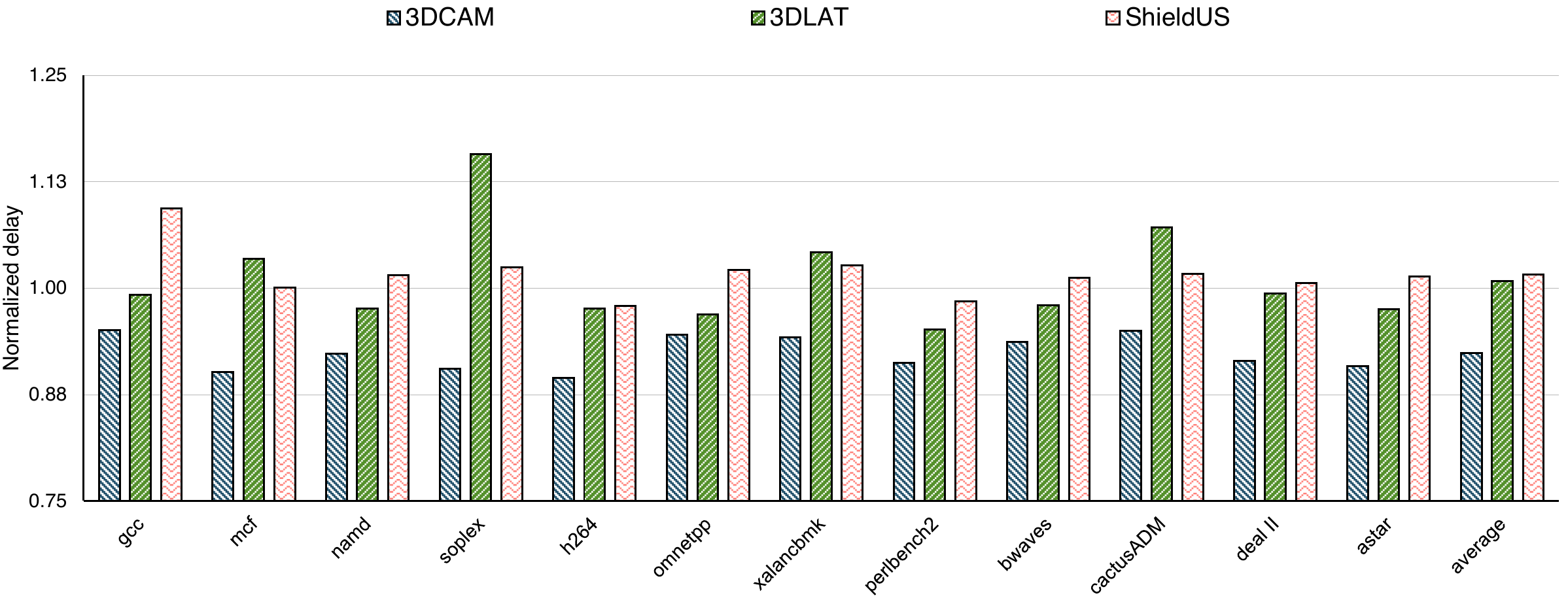}
	\centering
	\caption{Average transmission delay of 12 benchmarks which are normalized to the case that no crosstalk minimization technique is used. }
	\label{delay}
	\end{figure*}

Fig.~\ref{delay} demonstrates the transmission delay for several benchmarks from SPEC2006 suite. The delays are normalized to the case that no crosstalk minimization technique is used. As Fig.~\ref{delay} represents, 3DCAM can reduce the transmission delay of benchmarks by 9\% compared to the base uncoded case and in the best case 3DCAM could reduce transmission delay of \textit{soplex} benchmark by 25.7\% compared to 3DLAT method. Since 3DLAT (with $\omega = 4$) tries to code the input data in the manner that the coded output has no corsstalk class higher than 23C (based on our model), it can have a destructive effect on the transmissions time with lower crosstalk class. As the most transitions of \textit{soplex} benchmark fall into lower crosstalk classes, which is less than the average crosstalk class of 3DLAT coded outputs, 3DLAT even increases the transmission delay of this benchmark.
ShieldUS also could not effectively reduce the delay of experimented benchmarks. Inasmuch as this method can only reduce transmission delay of benchmarks with highly regular data and signals.
	
\subsection{Crosstalk Class Frequency Analysis}
\begin{figure}[!t]
\centering
\includegraphics[width=3.57in]{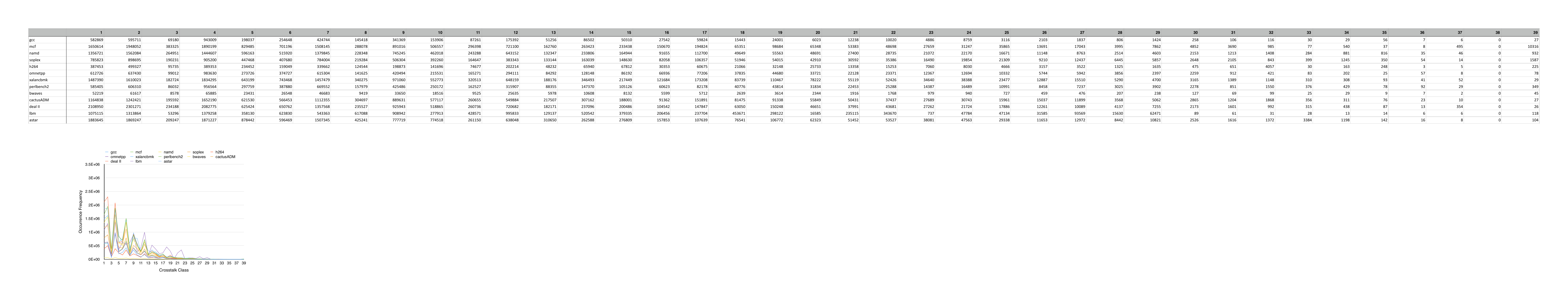}
\centering
\caption{The occurrence frequency of crosstalk classes before applying 3DCAM }
\label{class2}
\end{figure}

\begin{figure}[!t]
\centering
\includegraphics[width=3.57in]{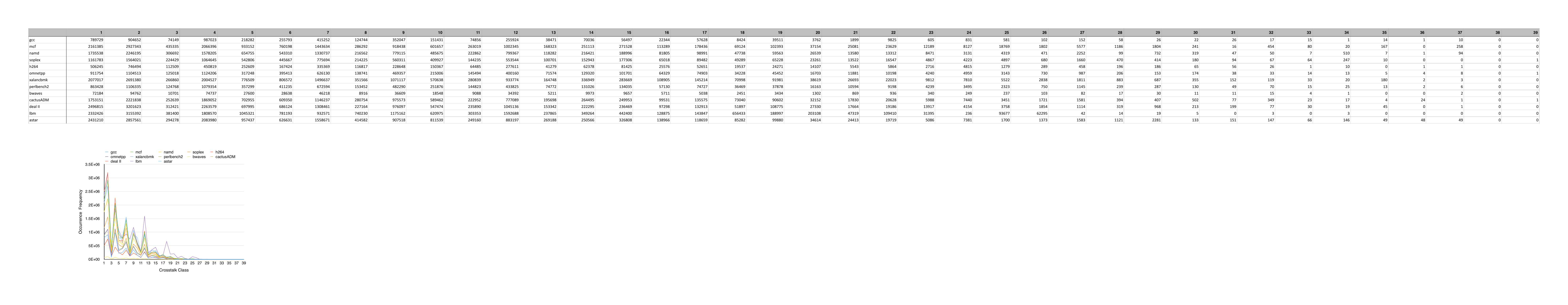}
\centering
\caption{The occurrence frequency of crosstalk classes after applying 3DCAM}
\label{class1}
\end{figure}

As the occurrence frequency of higher crosstalk classes directly affects the signal integrity and transmission time, the frequency of crosstalk classes is measured before and after applying the 3DCAM mechanism. Fig.~\ref{class2} and Fig.~\ref{class1} show the occurrence frequency of crosstalk classes before and after applying the 3DCAM mechanism, respectively. As these figures show, 3DCAM causes most of crosstalk patterns to fall into the left side of the chart. Namely it pushes the higher crosstalk classes to the lower crosstalk classes.

\subsection{Area Overhead}
As mentioned in Section.~\ref{proposed}, 3DCAM uses a simple coder which leads to less area overhead compared to all previous works.
Table \ref{table.3} demonstrates the coder and decoder area overhead of ShieldUS, 3DLAT, and 3DCAM mechanisms. For $3\times3$ TSV cluster in 3DLAT, the area of coder and decoder is 4264 $\mu m^2$ due to its large comparators. We estimate ShieldUS area overhead with the area of a $9\times9$ crossbar which is 218 $\mu m^2$. Finally, the area of 3DCAM is 116 $\mu m^2$.  These mechanisms are implemented and synthesized with 45nm technology using Synopsys Design Compiler.

\begin{table}[!h]
\renewcommand{\arraystretch}{1.5}
\caption{Area overhead of different mechanisms coder}
\label{table.4}
\centering
\begin{tabular}{| c| c|}
\hline
\textbf{Mechanism} & \textbf{Area} ($\mu$$m^2$) \\
\hline 
ShieldUS crossbar (only) & 218 \\
\hline
3DLAT & 4264 \\
\hline
3DCAM & 116 \\
\hline
\end{tabular}
\end{table}

\section{Conclusion}
In this paper, a crosstalk avoidance mechanism for TSV-to-TSV coupling capacitance is presented and a different crosstalk model has been discussed. The proposed mechanism decides about sending original data or retaining the TSV value in previous state based on a switch threshold (ST) parameter. 3DCAM reduces the frequency of higher crosstalk patterns which leads to reduce the interconnection delay. As compared to previous work, 3DCAM imposed negligible area overhead. GEM5 simulator is used to extract transitions of the data bus. We used real benchmarks taken from SPEC2006 suite. The simulation results showed that 3DCAM reduces transmission delay up to 25.7\%  while it reduces TSV overhead by 62.5\%  compared to 3DLAT mechanism. 
    
In our future work, first, we plan to consider a comprehensive crosstalk model based on capacitance and inductance coupling effects. Since the inductive coupling effect will increase in the near future, it has to be considered in conjunction with capacitance coupling effect. Second, presenting a probability model for each crosstalk classes in TSVs is another task for authors. Third, developing an analytical reliability model for TSV-to-TSV coupling effect in 3D ICs is  going to be discussed in the future work.

\section*{Acknowledgment}
The authors would like to thank the anonymous reviewers for their comments which were very helpful in improving the quality and presentation of this paper.

\bibliographystyle{abbrv}

\bibliography{references}

\end{document}